# Exploiting Parallelism on Shared Memory in the QED Particle-in-Cell Code PICADOR with Greedy Load Balancing

Iosif Meyerov[1], Sergei Bastrakov[2], Aleksei Bashinov[3], Evgeny Efimenko[3], Alexander Panov[1], Elena Panova[1], Igor Surmin[1], Valentin Volokitin[1], and Arkady Gonoskov[4,3,1]

[1] Lobachevsky State University of Nizhni Novgorod, Nizhni Novgorod, Russia
[2] Helmholtz-Zentrum Dresden-Rossendorf, Dresden, Germany
[3] Institute of Applied Physics, Russian Academy of Sciences, Nizhni Novgorod, Russia
[4] University of Gothenburg, Gothenburg, Sweden
`meerov@vmk.unn.ru`

**Abstract.** State-of-the-art numerical simulations of laser plasma by means of the Particle-in-Cell method are often extremely computationally intensive. Therefore there is a growing need for development of approaches for efficient utilization of resources of modern supercomputers. In this paper, we address the problem of a substantially non-uniform and dynamically varying distribution of macroparticles in a computational area in simulating quantum electrodynamic (QED) cascades. We propose and evaluate a load balancing scheme for shared memory systems, which allows subdividing individual cells of the computational domain into work portions with subsequent dynamic distribution of these portions between OpenMP threads. Computational experiments on 1D, 2D, and 3D QED simulations show that the proposed scheme outperforms the previously developed standard and custom schemes in the PICADOR code by 2.1 to 10 times when employing several Intel Cascade Lake CPUs.

## 1 Introduction

Numerical simulation of plasma by the Particle-in-Cell (PIC) method [1] for research in the use of high-intensity lasers is a rapidly developing area of computational physics. Interaction of intense laser pulses with different targets provides the possibility of exciting a complex collective electron dynamics in the plasma produced during this process. It opens up new opportunities for both research and solving important applied problems. One of the leading directions in this area is the study of effects of quantum electrodynamics (QED) in superstrong electromagnetic fields [2–6]. Large laser complexes planned for construction in the near future will allow creating electromagnetic fields necessary for the experimental observation of these processes. Today, an active study of future experiments is being carried out [6–9], with the cen-tral role being played by numerical simulation. Such simulations are performed with the Particle-in-Cell method extended with a module accounting for QED effects, a combination referred to as QED PIC.

QED PIC simulations are very computationally intensive and therefore are performed on supercomputers using highly optimized parallel software [10–18]. The problem of efficient implementation of the PIC method for parallel machines is quite well studied [19–27]. Fortunately, the method has a large potential for parallelization due to the local nature of the interactions. The spatial decomposition of the computational domain allows organizing parallel processing on distributed memory using MPI. Shared memory parallelization is usually done either by launching an MPI process on each core, or by using a combination of MPI and OpenMP. In this case, computationally intensive loops on particles or cells (supercells) are parallelized. This parallel processing scheme is widely used for plasma simulation codes. However, when modeling the QED effects, the problem of the explosive growth of the number of particles involved in the simulation resulting from the development of electromagnetic cascades comes to the fore. The exponential increase in the number of particles in a small area requires the development of special approaches to overcome the unacceptable expenditure of RAM and computational imbalance. The proper use of special procedures for thinning and merging particles (see, for example, [28]) allows us to control the memory usage but does not solve the problem of load balancing. Given the growing number of cores in modern CPUs, the problem of effectively parallelizing QED PIC codes for shared memory systems is becoming increasingly important.

Our previous study [29] compared five load balancing schemes for parallelizing a computational loop on cells containing substantially different numbers of particles. For problems with a relatively small workload imbalance, these schemes enabled achieving acceptable scaling efficiency. However, in case of a substantially non-uniform distribution, none of the considered schemes showed good results. As a unit of work all those schemes used processing particles in a cell. This is a natural choice for PIC codes using cell-based particle storage, it also simplifies parallel processing on shared memory. In this paper we propose a new, more sophisticated, scheme of parallel processing and load balancing. As a unit of work it uses processing all or some particles in a cell. It also handles the additional synchronization required when particles of the same cell are processed concurrently by different threads. Essentially, the new parallel processing scheme it is a generalization of the previously considered ones with some cells split into several portions of work. It increases the number of portions of work and allows avoiding the typical scenario of QED PIC simulations, when few cells with a huge number of particles limit scaling on shared memory. At the same time, the vast majority of cells remain as a single portion of work. We distinguish between the two types of cells at run time and only apply additional synchronization for the split cells, thus avoiding significant overheads.

The paper is organized as follows. In Section 2 we overview the QED PIC method. Section 3 introduces a baseline parallel algorithm. In Section 4 we give a detailed description of the new load balancing scheme. Section 5 presents numerical results. Section 6 concludes the paper.

## 2   An Overview of the Quantum Electrodynamics Particle-in-Cell Method

The Particle-in-Cell method [1] is commonly used to describe self-consistent dynamics of charged particles in electromagnetic fields. These entities are represented numerically as the main sets of data. Electromagnetic fields and current density are set on a grid, and particles are modeled as an ensemble of macroparticles with continuous coordinates. Each macroparticle in the simulation represents a cloud of physical particles of the same kind (with equal mass and charge) closely located in the coordinate space and with the same momentum. According to this duality of data representation, fields and particles are updated on different stages. There are four main PIC stages.

A field solver updates the grid values of the electromagnetic fields based on the Maxwell's equations. In PICADOR the conventional finite-difference time-domain Yee method based on the staggered grid is used [30]. On the particle motion stage the position and velocity of each particle are updated according to the Newton's law in the relativistic form, numerically integrated using an explicit method. Field interpolation from the grid to the particle positions is performed to compute the Lorenz force affecting the particles. Individual particle motion creates electric current, which is added to grid values on the current deposition stage, to be used in the field solver, thus completing the self-consistent system of equations. For efficiency, it is usually convenient to combine the stages of field interpolation, Lorenz force computation and integrating equations of particle motion into one stage, referred to as the particle push.

This standard PIC scheme can be extended by different modules which take into account various physical processes, such as ionization and collisions [12, 19, 31]. In the case of extremely strong electromagnetic fields QED processes come into play. Charged particles accelerated in extreme laser field may emit highly energetic photons, which in turn may decay into a pair of electron and positron [4, 32]. When repeatedly occurring, these processes may lead to an avalanche-like pair density growth, similar to avalanche gas breakdown, leading to development of the so-called QED cascades [33]. This dynamics may lead to formation of localized highly absorbing pair density distribution efficiently converting laser source energy into highly energetic gamma photons and charged particles, thus it can be treated as a source of antimatter (positrons), extremely dense electron-positron plasma, highly energetic electron bunches and photons [6, 7, 34, 35].

The QED cascade development is a complex process which depends not only on intensity, wavelength and polarization of the electromagnetic fields, but also on their structure. This makes theoretical analysis of laser-plasma dynamics very complex and nearly impossible when it comes to the highly non-linear stage of interaction. This makes computer simulation extremely useful for study of the plasma-field dynamics in complex field. At high laser intensity, when radiation reaction becomes essentially stochastic, it is important to consider single particle trajectories to describe plasma dynamics correctly. A PIC code by design relies on particles' representation as an ensemble of macroparticles and allows direct modeling of particle trajectories. Moreover, the extended PIC approach allows for treating high-energy photons as particles that are generated by electrons and positrons and can later decay into pairs [32, 33].

This approach utilizes dual treatment of the electromagnetic field: grid field values for the coherent low-frequency part and particles for the incoherent high-frequency part [18]. Photon emission and pair generation are probabilistic processes and their rates can be calculated under certain assumptions using expressions of QED based on the local field approximation [6]. Handling the considered QED events implies adding new particles associated with either emitted photons or produced particles.

During QED cascade development the number of physical particles may rapidly increase by many orders of magnitude, so the numerical scheme should be adapted for such conditions, preserving a reasonable number of macroparticles. This can be done by reweighting macroparticles in the scope of the thinout procedure. In this paper, we consider the PIC code PICADOR equipped with the adaptive event generator, which automatically locally subdivides the time step to account for several QED events in the case of high process rates. It employs a separate thinout module for each type of particles, which allows effective processing of the rapidly increasing number of particles. Methodological and algorithmical aspects of such an extended PIC scheme have been considered in [18]. Notably, a distribution of generated electron-positron plasma can be extremely localized due to the avalanche-like character of cascade development, its strong dependence on field intensity, and peculiarities of particles' motion. Particle processing in this case becomes a non-trivial problem due to a large workload imbalance, thus the technique of reducing the imbalance is of great interest.

## 3 Baseline Parallel Algorithm

Parallel processing in PICADOR is organized as follows. On distributed memory we use spatial domain decomposition of the simulation area, essentially standard for PIC codes of this kind. Each MPI process handles a subarea and stores all particles and grid values, with a few layers of ghost cells. Our previous work included load balancing on the level of distributed memory based on Cartesian rectilinear partitioning [24]. Inside each MPI process we employ parallelism with OpenMP, which is the focus of this paper. Notably, this shared-memory parallelism is largely independent of the distributed memory one and thus could be effectively considered separately.

The main computational workload of QED PIC simulations are particle operations: both particle-grid interactions in the core PIC algorithm and QED-specific processing. These operations are spatially local, which has important implications in terms of implementation. Firstly, particles need to be stored according to their positions so that processing nearby particles involves compact access to grid values. Two widely used strategies are periodical spatial sorting of particles, or storing them according to cells or supercells. Secondly, processing particles sufficiently far away from one another is completely independent and thus allows parallelization. It is particularly straightforward in case of cell- or supercell-based particle storage.

PICADOR follows this strategy. Particles are stored and processed separately for each cell, with each particle interacting only to closely located grid values. The radius of such interaction depends on the particle form factor and numerical schemes being used, but is constant throughout a particular simulation. It allows us to separate cells into subsets, so that particles in different cells of each subset do not affect each other and can therefore be processed in parallel without any synchronization. This scheme has been presented in our earlier work [29] and is illustrated in Fig. 1. We refer to processing each of these groups, as a 'walk'. Due to the subdivision, each loop over cells in a walk has independent iterations and synchronization has to be done only between walks. The minimum unit of workload is then processing particles of a cell.

We have investigated influence of load balancing schemes in the paper [29]. The first scheme parallelized the loop on cells in each walk using the standard OpenMP static schedule. Such a scheme worked excellent only in relatively balanced scenarios. The second scheme used OpenMP dynamic scheduling. This scheme substantially reduced imbalance, but led to large balancing overhead. In the other three schemes, the cells in each walk were occasionally sorted by decreasing the number of particles in order to improve load balancing potential. The third scheme then used OpenMP dynamic scheduling, the fourth scheme manually distributed the cells into threads, using a greedy strategy and avoiding the overhead of dynamic balancing of OpenMP. The fifth scheme at times used the OpenMP dynamic schedule, saved the distribution obtained and used it for several subsequent iterations over time. In this way, effective load distribution was achieved for systems with relatively slow dynamics.

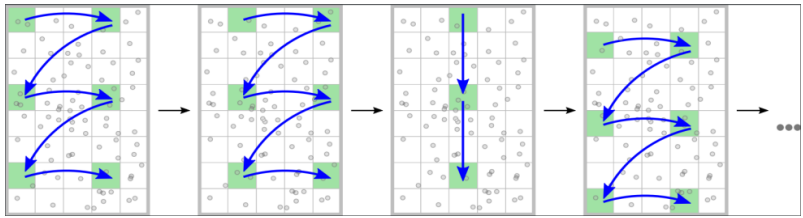

**Fig. 1.** An example of cells split into four walks. Particles are represented with the grey dots. Cells inside each walk are processed independently in parallel. Walks are performed sequentially with a barrier between walks.

For many PIC simulation scenarios, particle distribution changes rather slowly relative to the cell size, thus the standard OpenMP static schedule or one of the custom schemes provide excellent load balancing. For QED PIC, however, some cells can have significantly more particles than others and the distribution of particles in the simulation area can vary significantly over time. In such a case a cell is too coarse of a workload unit. Therefore we developed a new load balancing scheme employing cell subdivision, which is described in the following section.

## 4 Dynamic Load Balancing Scheme

The main idea of the new scheme is to treat subsets of particles in a cell as separate pieces of work. This allows balancing the workload so that each thread processes almost the same number of particles. An illustration of the algorithm is given in Fig. 2, a more detailed description is presented below.

Compared to the schemes described in the previous section, the walks play the same role, but now processing a cell can consist of several tasks, each handling a subset of particles. Importantly, tasks of the same cell are dependent, but tasks of different cells are not. Each thread has a queue of tasks in which no more than one task corresponds to a subdivided cell. Thus, a relatively small number of cells not exceeding the number of threads can be subdivided.

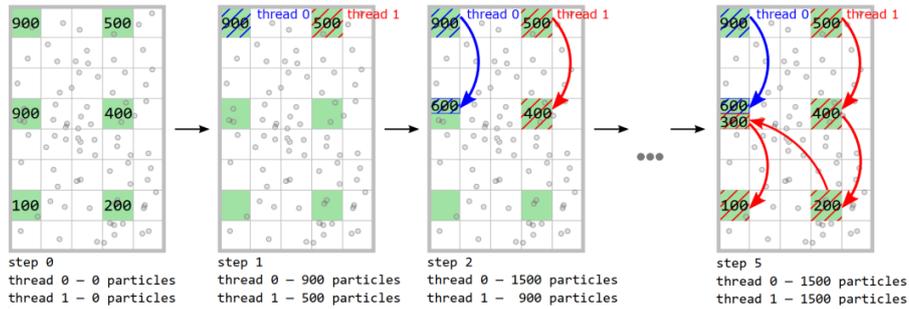

**Fig. 2.** An example of the new load balancing scheme applied to a single walk. The numbers represent the amounts of particles in cells. The blue and red arrows illustrate two threads working in parallel. One of the cells is subdivided into two tasks.

The pseudocode of this scheme is as follows:

```
1.   updateField(Grid);
2.   for Walk in Walker:
3.     createTasksQueue();
4.   #pragma omp parallel:
5.     for Task in TasksQueue[threadIndex]:
6.       for Particle in Task:
7.         process(Particle);
8.   for walk in walker:
9.   #pragma omp parallel for:
10.    for Task in TasksQueue[threadIndex]:
11.      for Particle in Task:
12.        currentDeposition(Particle);
```

We use the following greedy algorithm, which is linear-time in terms of number of cells. At the beginning, an empty queue is created for each thread. Further, the current cell with particles is completely added to the tasks queue of the thread if the total size of the tasks in the queue does not violate the ideal balance by more than $M$ times ($M$ is a parameter of the algorithm). Otherwise, the cell is divided into two parts so that the including of the first batch of work to the queue of the current thread corresponds to the ideal balance. Then, the tasks queues of other threads are computed in a similar fashion.

The main implementation challenge is avoiding time-consuming synchronizations caused by the subdivision of cells and their distribution between threads. For exam-ple, when processing the movement of particles between cells, the algorithm remembers the numbers of the corresponding particles and processes them last so that each cell is processed by only one thread. Similarly, atomic operations have to be used at the current deposition stage. However, experiments have shown that these complications do not lead to significant overhead, whereas the scheme substantially speeds up the QED PIC simulations.

## 5   Numerical Results

First, we analyze the effectiveness of the load balancing schemes for a two-dimensional test problem. We use a $160 \times 160$ grid and $2.56 \times 10^6$ particles, 100 particles per cell on average. Initial sampling of particles is done with the normal distribution with the mean in the center of the simulation area and a diagonal covariance matrix with the same variance for both variables (the spatial steps are also the same). We consider three variance values: $\sigma_1^2 = 25\Delta x/8, \sigma_2^2 = 2\sigma_1^2, \sigma_3^2 = 3\sigma_1^2$. The smallest variance $\sigma_1^2$ corresponds to the most non-uniform distribution. Simulations were performed for 1000 time steps without the QED effects. Computations were carried out on a node of a supercomputer with the following parameters: $2 \times$ Intel Xeon Gold 6132 (28 cores overall), 192 GB RAM. We measured the computation time using 5 load balancing schemes implemented in [29] and the new scheme. Experiments have shown that for the first (most unbalanced) problem, the new scheme outperforms the best of the others by factor of 4.4, the corresponding speedups for the second and third problems are 1.9 and 1.1.

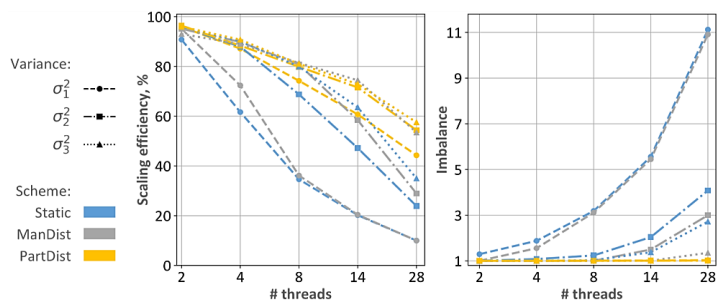

**Fig. 3.** Scaling efficiency and workload imbalance in 3 test problems with different values of variance when employing 3 load balancing schemes on the 28-core CPU.

Fig. 3 shows the scaling efficiency and the imbalance of the computational load when using from 1 to 28 cores when solving 3 test problems. Same as in [29], we estimate workload imbalance as $I = \max_w \left( \underset{i \in \{0,\ldots,N-1\}}{\mathrm{mean}} \left( \frac{\max_t P_{wti}}{\mathrm{mean}_t P_{wti}} \right) \right)$, where $P_{wti}$ is a number of particles processed by the thread $t$ within walk $w$ on $i$-th out of total $N$ iterations. In order not to overload the figure, only the data for static balancing ('Static'), the best of the previously introduced dynamic schemes ('ManDist') and the new scheme ('PartDist'), are presented. It is shown that in the substantially unbalanced task, the new scheme reduces the imbalance and improves the scaling efficiency.

Secondly, we study performance of the load balancing schemes in state-of-the-art simulations that take into account the QED effects. In these applications, the processes become much more complicated, since the concentration of particles in the local regions of the computational area is not only highly unbalanced, but also changes substantially over time. It is particularly interesting to estimate the gain from the use of the new dynamic scheme in such scenarios.

We consider a highly unbalanced problem of the QED cascade development in extreme laser fields described in detail in [36]. According to recent studies [8, 32, 35, 37] the more preferable field structure is counterpropagating laser beams. We would like to pay particular attention to the case of circular polarization, as one of the fundamental cases. In a wide range of intensities of the circularly polarized field highly localized electron-positron structures can be formed in the vicinities of electric field nodes, antinodes or in both types of regions [36].

The maximum intensity of each of counter-propagating pulses is chosen to be $I_0 = 10^{25}\ W/cm^2$, which can be obtained on planned $100\ PW$ laser systems. The wavelength is $0.8 \mu m$. For the sake of simplicity we consider half infinite pulses with a 1 wave period front edge. An electron-positron plasma slab with width of one wavelength and density $1 cm^{-3}$ serves as a seed and is located at the center of the simulation area. Incident laser pulses compress seed plasma. Laser pulses overlap, standing wave is formed and a QED cascade starts to develop. At the considered intensity during its development plasma is highly localized in the vicinity of the antinode.

All experiments were performed on the Intel Endeavour supercomputer with high-end CPUs of the Cascade Lake generation. Performance results were collected on the following cluster nodes: 2 × Intel Xeon Platinum 8260L CPU (48 cores overall), 192 GB of RAM. In all runs we employed 1 MPI process per socket, 2 OpenMP threads per core. The code was built using the Intel Parallel Studio XE software package.

First, we consider the 1D case where the number of cells most populated by particles is minimal. The simulation box is $2 \mu m$ long and the number of cells is 128. The time step is equal to $1/200$ of the laser period. The initial number of particles and the threshold of particle thinning are $10^6$ and $2 \times 10^6$, respectively. In the 1D simulation we employ 1 node of the supercomputer (48 cores).

Experiments show that computation time of the static scheme and two most flexible of the old dynamic schemes are roughly the same, while the new scheme is better by almost an order of magnitude. This is due to the fact that the new scheme balances the workload much better by subdividing the cells. Fig. 4 shows how imbalance of the computational load changes over time. It is calculated for every chunk of 100 consecutive iterations. The imbalance in Fig. 4 on the left is calculated using the profiler based on computation time measurements. The imbalance in Fig. 4 on the right is estimated by the number of particles processed by the threads. Given that for the 2D and 3D simulations we obtained similar results, we can conclude that the employed model of particle imbalance works well in the case of modeling of the QED cascades.

In 2D and 3D simulations the box is $2 \mu m \times 8 \mu m (\times 8 \mu m)$, the number of cells is $64 \times 112 (\times 112)$, the initial number of macroparticles and the threshold of particle thinning are $5 \times 10^6$ in the 2D case and $2.5 \times 10^6$ in the 3D case. We consider Gaussian beams with 1 wavelength waist radius. The time step is the same as in the 1D case. The 2D and 3D cases are much more computationally intensive, therefore we employ 2 nodes of the supercomputer (4 CPUs, 96 cores overall). The PICADOR code has load balancing schemes for clusters, but in the considered case, workloads at different nodes are balanced due to the symmetry of the problem. In contrast, uniformly distributing the work among dozens of cores on a single node is problematic.

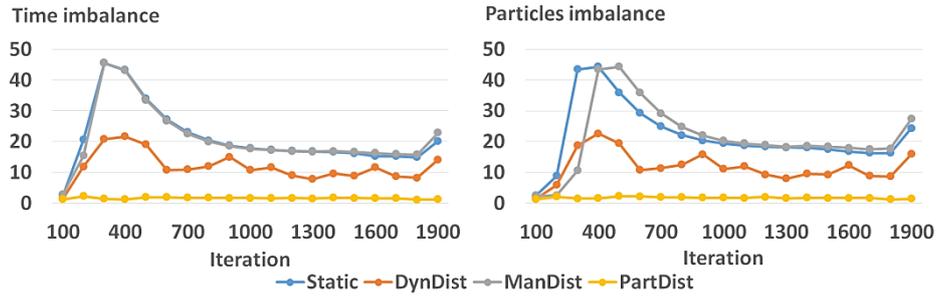

**Fig. 4.** 'Time imbalance' vs. 'Particles imbalance' in the 1D simulation of the QED cascades.

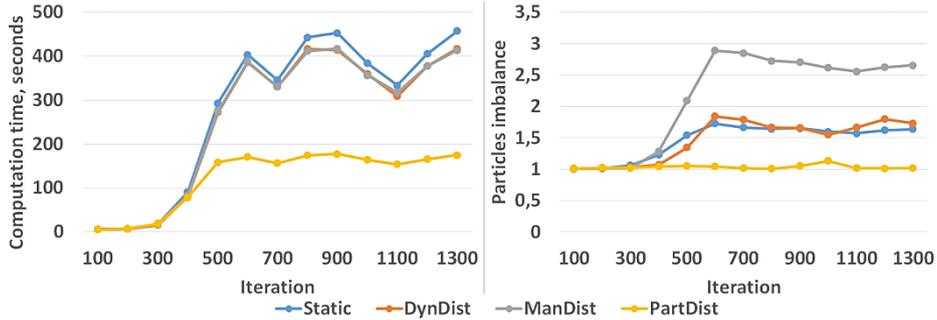

**Fig. 5.** Computation time of different load balancing schemes in the 3D simulation of the QED cascades. The new scheme ('PartDist') outperforms other schemes.

Fig. 5 shows how different load balancing schemes work for the considered 3D simulation. In the first 300 iterations, there is no imbalance (Fig. 5, on the right) and all the schemes work for approximately the same time (Fig. 5, on the left). Further, the electromagnetic cascade begins to develop and the new scheme, unlike the others, allows keeping the imbalance under control. Similar behavior is observed in the 2D simulation. Finally, when calculating 1300 iterations, the new scheme speeds up the simulation by 2.5 times in the 2D problem and by 2.1 times in the 3D problem.

## 6   Conclusion

In the paper, we addressed the problem of the inefficient utilization of new CPUs with a large number of cores in PIC laser plasma simulations. We concentrated on simula-

10tions including the development of electromagnetic cascades, which often led to a non-uniform and time-varying concentration of particles in the computational domain. To overcome the load imbalance, we developed and implemented a special scheme in the PICADOR code that allows subdividing cells with a large number of particles. This approach substantially increased the potential for parallelization. The scheme was tested on several QED simulations. The results showed that, in the absence of a significant imbalance, the new scheme performs approximately the same as the schemes previously developed in PICADOR, whereas in other cases it is ahead of them by a factor of 2.1 to 10, depending on the simulation.

Note that the current results were obtained for the problems whose internal symmetry allows using several cluster nodes in a straightforward way, or relying on the rectilinear load balancing scheme (see [24] and references therein for details). However, if a problem requires the use of dozens of nodes, there is a need for a smart combination of balancing schemes on shared and distributed memory. Development and analysis of such schemes is one of the directions of further work.

**Acknowledgements.** The work is supported by the RFBR project No. 18-47-520001.# 7      References

1. Dawson, J.: Particle simulation of plasmas, Rev. Mod. Phys. 55, 403 (1983).
2. Elkina, N.V. et al.: QED cascades induced by circularly polarized laser fields, Phys. Rev. Spec. Top.–Accel. Beams 14, 054401 (2011).
3. Sokolov, I. et al.: Numerical modeling of radiation-dominated and quantum electrodynamically strong regimes of laser-plasma interaction, Phys. Plasmas 18, 093109 (2011).
4. Ridgers, C.P. et al.: Modelling gamma-ray photon emission and pair production in highintensity laser-matter interactions, J. Comput. Phys. 260, 273 (2014).
5. Grismayer, T. et al.: Laser absorption via quantum electrodynamics cascades in counter propagating laser pulses, Physics of Plasmas, Vol. 23. P. 056706 (2016).
6. Gonoskov, A. et al.: Ultrabright GeV Photon Source via Controlled Electromagnetic Cascades in Laser-Dipole Waves, Physical Review X, Vol. 7. P. 041003 (2017).
7. Efimenko, E.S. et al: Laser-driven plasma pinching in e−e+ cascade. Phys. Rev. E 99, 031201(R) (2019).
8. Tamburini, M., Di Piazza, A., Keitel, C. H.: Laser-pulse-shape control of seeded QED cascades, Scientific reports, 7(1), 5694 (2017).
9. Samsonov, A. S., Nerush, E. N., Kostyukov, I. Y.: QED cascade in a plane electromagnetic wave, arXiv preprint arXiv:1809.06115 (2018).
10. Brady, C. S., Arber, T. D.: An ion acceleration mechanism in laser illuminated targets with internal electron density structure, Plasma Phys. and Contr. Fusion, 53(1), 015001 (2011).
11. Fonseca, R. A. et al.: OSIRIS: A three-dimensional, fully relativistic particle in cell code for modeling plasma based accelerators. In ICCS, pp. 342-351, Springer, Berlin (2002).
12. Bussmann, M. et al.: Radiative Signatures of the Relativistic Kelvin-Helmholtz Instability. In SC'13. ACM, New York (2013).
13. Derouillat, J. et al.: Smilei: a collaborative, open-source, multi-purpose particle-in-cell code for plasma simulation. Computer Physics Communications, 222, 351-373 (2018).
14. Pukhov, A. Three-dimensional electromagnetic relativistic particle-in-cell code VLPL (Virtual Laser Plasma Lab). Journal of Plasma Physics, 61(3), 425-433 (1999).

15. Bowers, K.J. et al: Ultrahigh performance three-dimensional electromagnetic relativistic kinetic plasma simulation. Phys. Plasmas. 15 (5), 055703 (2008).
16. Friedman, A. et al.: Computational methods in the warp code framework for kinetic simulations of particle beams and plasmas, IEEE Trans. on Pl. Sci., 42(5), 1321-1334 (2014).
17. Surmin, I. A., et al.: Particle-in-Cell laser-plasma simulation on Xeon Phi coprocessors. Computer Physics Communications, 202, 204-210 (2016).
18. Gonoskov, A., at al.: Extended Particle-in-Cell schemes for physics in ultrastrong laser fields: review and developments, Phys. Rev. E, 92, 023305 (2015).
19. Fonseca, R.A.: Exploiting multi-scale parallelism for large scale numerical modelling of laser wakefield accelerators, Plasma Phys. Control. Fusion. 55 (12), 124011 (2013).
20. Decyk, V.K., Singh, T.V.: Particle-in-cell algorithms for emerging computer architectures, Comput. Phys. Commun. 185 (3), 708-719 (2014).
21. Germaschewski, K. et al.: The Plasma Simulation Code: A modern particle-in-cell code with patch-based load-balancing. J. Comp. Phys., 318 (1), 305–326 (2016).
22. Beck, A. et al.: Load management strategy for Particle-In-Cell simulations in high energy physics. Nucl. Inst. Meth. in Phys. Res. A, 829 (1), 418–421 (2016).
23. Vay, J.-L., Haber, I., Godfrey, B.B. A domain decomposition method for pseudo-spectral electromagnetic simulations of plasmas, J. Comp. Phys. 243 (15), 260–268 (2013).
24. Surmin, I. et al.: Dynamic load balancing based on rectilinear partitioning in particle-in-cell plasma simulation, In: Malyshkin, V. (ed.) PaCT 2015. LNCS, vol. 9251, pp. 107–119. Springer, Cham (2015).
25. Kraeva, M.A., Malyshkin, V.E.: Assembly technology for parallel realization of numerical models on MIMD-multicomputers. Future Gener. Comput. Syst. 17, 755–765 (2001)
26. Vshivkov, V.A., Kraeva, M.A., Malyshkin, V.E.: Parallel implementations of the particle-in-cell method, Programming, No. 2, 39–51 (1997).
27. Surmin, I. et al.: Co-design of a particle-in-cell plasma simulation code for Intel Xeon Phi: a first look at Knights Landing, In ICA3PP. LNCS, pp. 319-329, Springer, Cham. (2016).
28. Vranic, M., Grismayer, T., Martins, J.L., Fonseca, R.A., Silva, L.O.: Particle merging algorithm for PIC codes, Computer Physics Communications, 191:65–73 (2015).
29. Larin, A., et al.: Load Balancing for Particle-in-Cell Plasma Simulation on Multicore Systems. In: Wyrzykowski R., Dongarra J., Deelman E., Karczewski K. (eds) PPAM 2017. LNCS, vol 10777. Springer, Cham (2018).
30. Taflove, A., Hagness, S.: Computational Electrodynamics: The Finite-Difference Time-Domain Method, The Artech House Ant. and Prop. Lib. (Art. House, Inc., Boston) (2005).
31. Vay, J.-L., et al.: Simulating relativistic beam and plasma systems using an optimal boosted frame. J. Phys. Conf. Ser. 180 (1), 012006 (2009).
32. Nerush, E.N., et al.: Laser Field Absorption in Self-Generated Electron-Positron Pair Plasma. Phys. Rev. Lett. 106, 035001 (2011).
33. Bell, A.R., Kirk, J.G.: Possibility of Prolific Pair Production with High-Power Lasers. Phys. Rev. Lett. 101, 200403 (2008).
34. Vranic, M., Grismayer, T., Fonseca R.A., Silva, L.O.: Electron-positron cascades in multiple-laser optical traps. Plasma Phys. Control. Fusion 59, 014040 (2016).
35. Efimenko E.S., et al.: Extreme plasma states in laser-governed vacuum breakdown. Sci. Rep. 8, 2329 (2018).
36. Bashinov, A.V., et al.: Particle dynamics and spatial e−e+ density structures at QED cascading in circularly polarized standing waves, Phys. Rev. A 95, 042127 (2017)
37. Jirka, M. at al.: Electron dynamics and $\gamma$ and $e-e+$ production by colliding laser pulses, Phys. Rev. E 93, 023207 (2016).